# Electronic properties of the Rutile-type dioxide SnO2 material doped by sulfur element: DFT study

S. IDRISSI[1,*], L. BAHMAD[1,*] and A. BENYOUSSEF[2]

[1] Laboratoire de la Matière Condensée et des Sciences Interdisciplinaires (LaMCScI), Mohammed V University of Rabat, Faculty of Sciences, B.P. 1014 Rabat, Morocco.

[2] Hassan II Academy of Science and Technology, Av. Mohammed VI, Rabat, Morocco.

**Abstract:**

In this work, we study the effect of doping the Rutile-type dioxide SnO2 material by the non-metal Sulfur (S) atoms on the electronic properties. In fact, we have used the ab-initio method applied on the basis of the Density Functional Theory (DFT) using the Quantum Espresso code. Through the density of states and the band structure calculations for different concentrations have been deduced.

When doping the SnO2 material with 6% of Sulfur (S), we found a perfect symmetry between up and down spin states in the total DOS confirming the non-magnetic behavior of this material doped SnO2 with 6% of Sulfur. It is also worth to note that the SnO2 material doped with 6% of Sulfur, exhibits a semiconductor of the P-type. Moreover the band gap decreases when increasing the concentration of doping the SnO2 by Sulfur. Our results are in good agreement with the existing literature both experimental and theoretical.

**Keywords:** Rutile-Type dioxide; SnO2; Sulfur element doping; Electronic properties; DFT method; Quantum Espresso.

[*]) Corresponding authors: samiraidrissi2013@gmail.com (S. I); bahmad@fsr.ac.ma (L.B);

I.  Introduction

Lately different scientists, Metal Oxide Semiconductors (MOS) like SnO2, CuO, ZnO, NiO and TiO2 are thought and created because of their selective properties and torpid applications to be specific optoelectronics, Nano electronics, storage devices, photonic de-indecencies, catalysis and spintronic applications [1-9].

The Rutile-type dioxide SnO2 material is an important semiconductor with wide-band gap (3.6 eV at 300 K) and presents spectacular behavior such as high carrier density, high transparency, chemical and thermal stability [10, 11]. Furthermore, which are utilized in endless applications, for example, level board shows, photo detector, solar cells, defensive covering, fluid precious stone presentation, straightforward conductive terminal, attractive capacity media, battery-powered Li-particle batteries and sensors [12, 13]. Then again, the SnO2 material has arranged principally with various morphologies, for example, nano-flowers [14–16], nanorods [17–21], nanofibres [22, 23], nano sheets [24], nanoparticles [25–27] and slender movies [28] just as blending the material in with respectable metals and their oxides, for example, Au [29–31], Ag [32–34] and Pd [35, 36].

On the other hand, Coey et *al*. [37] proposed a trade component that in volves the Vac(O)s named F- center, and they expected that TM–Vac(O)– TM gatherings would be basic in Fe-doped SnO2. Another principle disputable issue is that doping nonmagnetic. The SnO2 doped with Cu and Ni utilizing the co-precipitation strategy has been examined with hardly any reports accessible. In addition, P. Pascariu (Dorneanu) et *al*. [38]. As of late the unadulterated SnO2 NPs, Cu and Ni doped SnO2 NPs by co-precipitation technique and auxiliary, morphological, optical, attractive and electrical properties were studied [39].

In the present work, we study and analyze the structural and electronic properties of the Rutile-Type dioxide pure and doped by different concentrations of the sulfur element. In fact, the effect of doping this material has been interpreted, as function of the concentrations: 6 %, 13 %, 20%, 26% and 33% of sulfur. For this purpose, we study such properties by using the DFT method under the Quantum Espresso package. Indeed, some of our recent works have been based on such simulations applying the DFT method and other numerical simulations [40-44].

This paper is organized as follows: In section II, we describe the method and calculations. In section III, we present and discuss the results of ab-initio calculation of the pure and the doped SnO2 material. We provide our conclusions in the section IV.

## II. Method and calculations

The results found in this paper are investigated on doping effect on the electronic properties of the SnO2 material, using the Quantum Espresso code [50] on the basis of the Density Functional Theory (DFT).

The obtained results and calculations are simulated under norm-conserving pseudopotentials [51] for calculations without spin–orbital coupling (SOC) calculations. The exchange–correlation functional is treated using generalized-gradient approximation (GGA) of Perdew–Burke–Eruzerhof (PBE) [52].

In the study for the pure SnO2 material, the valence states of Sn and O elements are $5s^25p^2$ and $2s^22p^4$. For the nonmetals doping, the valence states of: S, is: $3s^23p^4$.

To ensure sufficiently accurate total energy calculations, a plane-wave basis set cut-off energy of 640.20 eV. The special points sampling integration in the Brillouin zone were employed using the Monkhorst-Pack method [53], with 5×5×5 special k-points mesh. In our simulations, the convergence threshold for self-consistent-field (SCF) iteration was set at $10^{-6}$ eV. For the density of states (DOS) for each perovskite, the width of the Gaussian smearing was 0.1 eV.

## III. Results of the Ab-initio method

### a. Structural properties

The unit cell of rutile SnO2 is a simple tetragonal lattice with a basis of two formula units. The unit cell of SnO2 is characterized by the two lattice constants a, c, and the internal parameter u. The coordinates of these basis can be taken as cations at (0, 0, 0) and (1/2, 1/2, 1/2) and anions at ± (u, u, 0) and ±(u, -u, 1/2), with u=0.307 [54]. This material belongs to the space group P42/mnm N° 136 with the face-centered cubic phase. A Schematic representation of the pure SnO2 material is illustrated in Fig.1. We have reproduced this plot by using the Vesta package [55].

The total energy of the pure SnO2 is presented in Fig.2 (a) as a function of the lattice parameter a (Å) and as a function of the lattice parameter c (Å) in Fig.2 (b). From Fig.2 (a) the total energy is reaches it's minimum for a=4.72 Å. While the value of the parameter c minimizing this energy is c= 3.17 Å. Moreover, the table 1 illustrates a comparison between our results and existing literature of the lattice parameters a (Å), b (Å) and c (Å) of the pure SnO2 material. It is clear that the obtained results in this work are close both the experimental [12, 27, 39] and theoretical [56-59] values given in the literature.

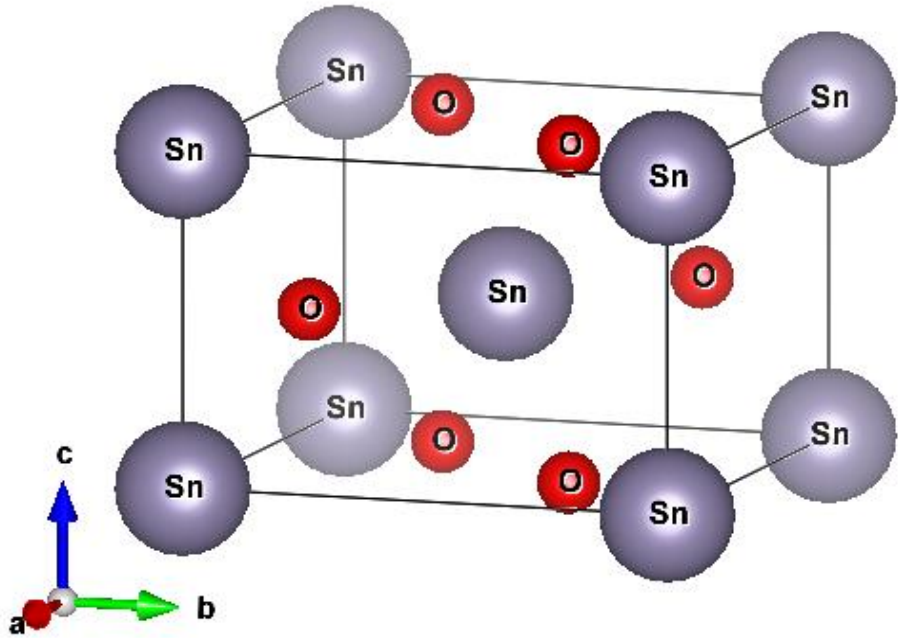

*Fig.1: A Schematic representation of the SnO2 using the Vesta package [55], the oxygen atoms are represented by red spheres, while the blue spheres correspond to the Sn atoms.*

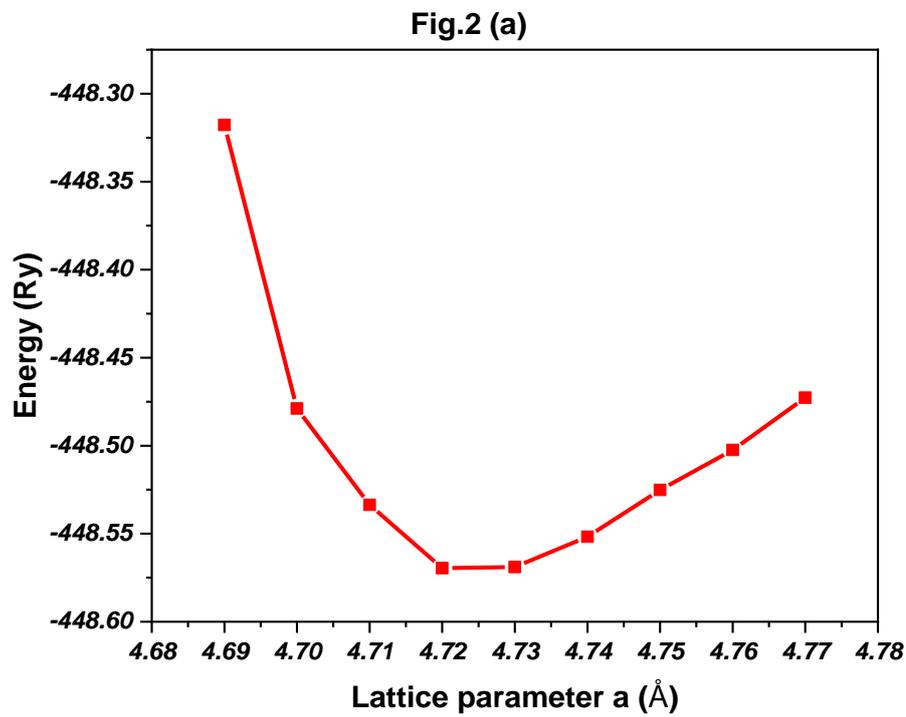

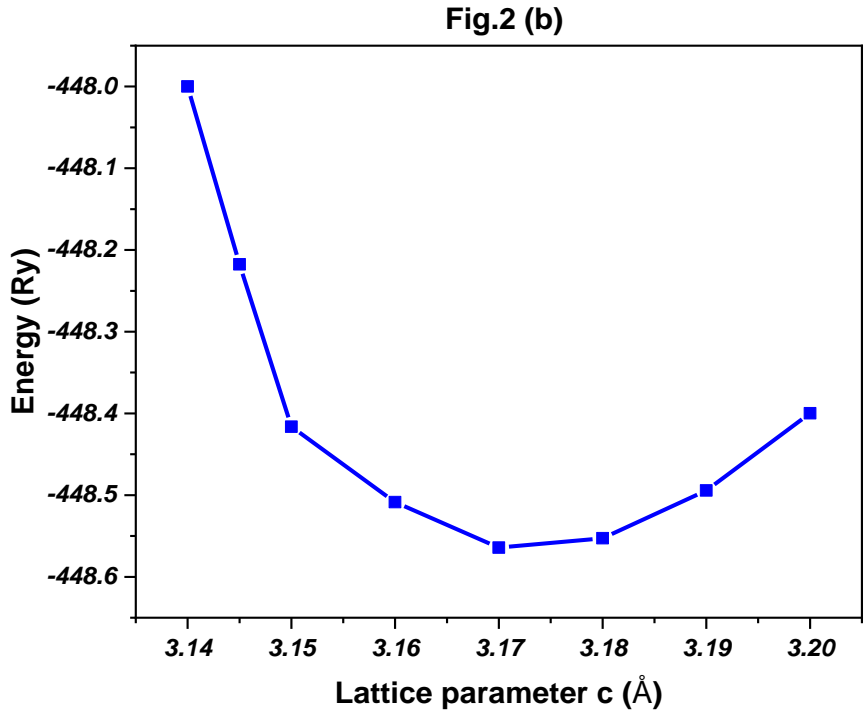

*Fig.2: The total energy of the pure SnO2 material, in (a) as a function of the lattice parameter a (Å) and b (Å), in (b) as a function of the lattice parameter c (Å).*

|  | Cell parameters (Å) | |
| --- | --- | --- |
|  | a=b | c |
| Present work | 4.72 | 3.17 |
| Experimental studies | 4.707 [39] | 3.196 [39] |
|  | 4.743 [12] | 3.198 [12] |
|  | 4.735 [27] | 3.188 [27] |
| Theoretical studies | 4.712 [56] | 3.174 [56] |
|  | 4.580 [57] | 3.080 [57] |
|  | 4.715 [58] | 3.194 [58] |
|  | 4. 673 [59] | 3.149[59] |

*Table 1: The lattice parameters: a (Å), b (Å) and c (Å) of the pure SnO2 material.*

### b. Electronic properties of the pure SnO2 material

In order to study the electronic properties of the pure SnO2 compound, we have presented the obtained results of the density of states (DOS) in Fig.3 (a) and the band structure of this material in Fig.3 (b). In fact, the DOS of the pure SnO2 material presented in Fig.3 (a) shows that this material is a semiconductor of N-type. Moreover, the perfect symmetry between up and down spin states confirms the non-magnetic behavior of this material. In addition, the O-2p orbital is the most contributing in the total DOS. On the other hand, the band structure of the pure SnO2 material presented in Fig.3 (b) exhibits a band gap value of about 2.5 eV. This band gap is found to be direct at the Γ point. It is worth to note that our results are a good agreement with the experimental values: 3.6 eV given in Ref [63] and 2.9 eV found in the Ref [64]. Some other theoretical values of this parameter are: 1.8 eV see Ref [60], 2.2 eV see Ref [61] and 2.76 eV see Ref [62].

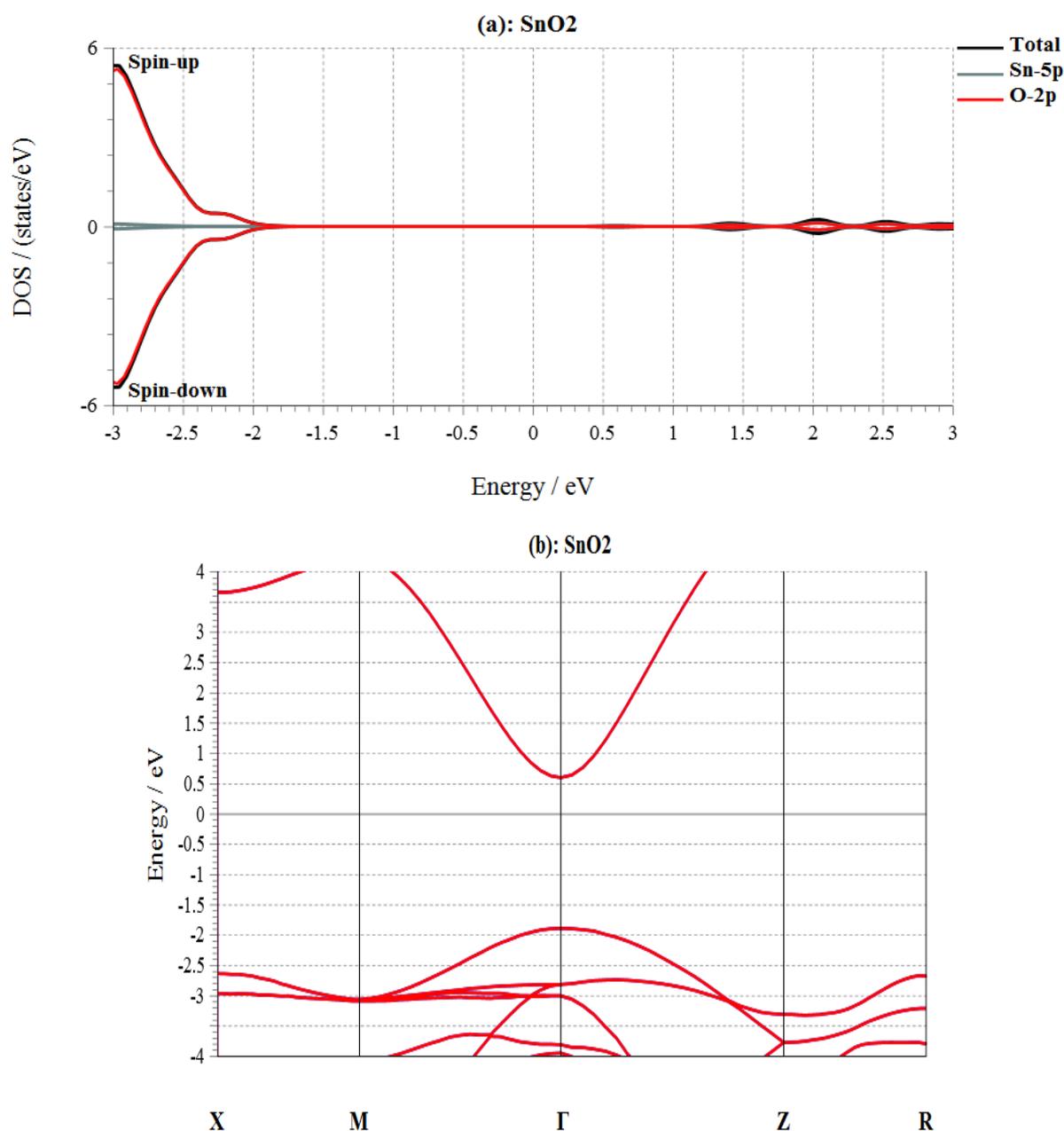

*Fig.3: Total and partial density of states (DOS) in (a), band structure in (b), of the pure SnO2 material.*

**c. Electronic properties of the SnO2 material doped with the sulfur element**

When doping the SnO2 material with 6% of Sulfur (S), we illustrate the obtained results in Fig.4. Such figure presents the total and partial density of states (DOS) in Fig. 4 (a) and the band structure in Fig.4 (b). A perfect symmetry is observed in the total DOS confirming the non-magnetic behavior of the doped SnO2 material with 6% of Sulfur. Also, the O-2p and S-2p orbitals are dominate in the valence band. While the Sn-5p orbital is the most contributing in the conduction band of total DOS. It is also worth to note that the SnO2 material doped with 6% of Sulfur, exhibit a semiconductor of the P-type. On the other hand, band gap corresponding to the doping SnO2 material with 6% of Sulfur decreases from 2.5 eV for the pure case to become 2.3 eV. The direct band gap of the pure SnO2 case has changed to in indirect band gap when doping SnO2 with 6% of Sulfur (S). The band gap value decrease to reach 1.8 eV when doping with 13 % of Sulfur, see Figs.5 (a) and 5 (b). When doping the SnO2 material doped with 13% of Sulfur, display a semiconductor of the N-type.

The Fig.6 summarizes the results of doping the SnO2 material with 20 % of Sulfur. The total and partial DOS are presented in Fig.6 (a). The band structure is presented in Fig.6 (b) showing a direct band gap at Γ point with the value 1.5 eV. The doping with 26% of Sulfur (S) keeps the material in the N-type semiconductor. When doping with 26% of Sulfur, we provide in Fig.7 (a) the total and partial density of states of this material. Such figure shows a perfect symmetry between up and down spin states, with a N-type semiconductor behavior. The band structure is given in Fig.7 (b) showing a direct band gap with the value 1.25 eV. By increasing the doping ratio to 33% of Sulfur, we illustrate in Figs.8 (a) and 8 (b) the total and partial density of states and corresponding band structure, respectively. Such concentration of doping the SnO2 material leads to an intrinsic semiconductor character with a direct gap at Γ point corresponding the value: 1.05 eV. The Fig.9 summarizes the variation of the total energy and the band gap of the doped SnO2 material as a function of the concentration (%) of impurities of sulfur. From this figure, it is found that the total energy increases, while the band gap decreases when increasing the concentration of doping. The table 2 regroups the different values of the band gap of the pure and several concentration of the SnO2 doped material by sulfur. Such table compares also our results with the different values existing in the literature. Our results are in good agreement with the similarity behavior in the literature. In fact, J. Divya *et al*.[39]. provide the band gap energy values: 2.84 eV for the pure SnO2 material, 2.50 eV for the Cu doped

SnO2 material and 2.30 eV for the Ni-doped SnO2 material, respectively. In addition, L. Soussi, *et al.* [65]. found the band energy values: 3.8 eV for the pure SnO2 material, 3.5 eV for the Co doped SnO2 material, 3.4 eV for the Fe-doped SnO2 material and 3.2 eV for the Ni-doped SnO2 material, respectively.

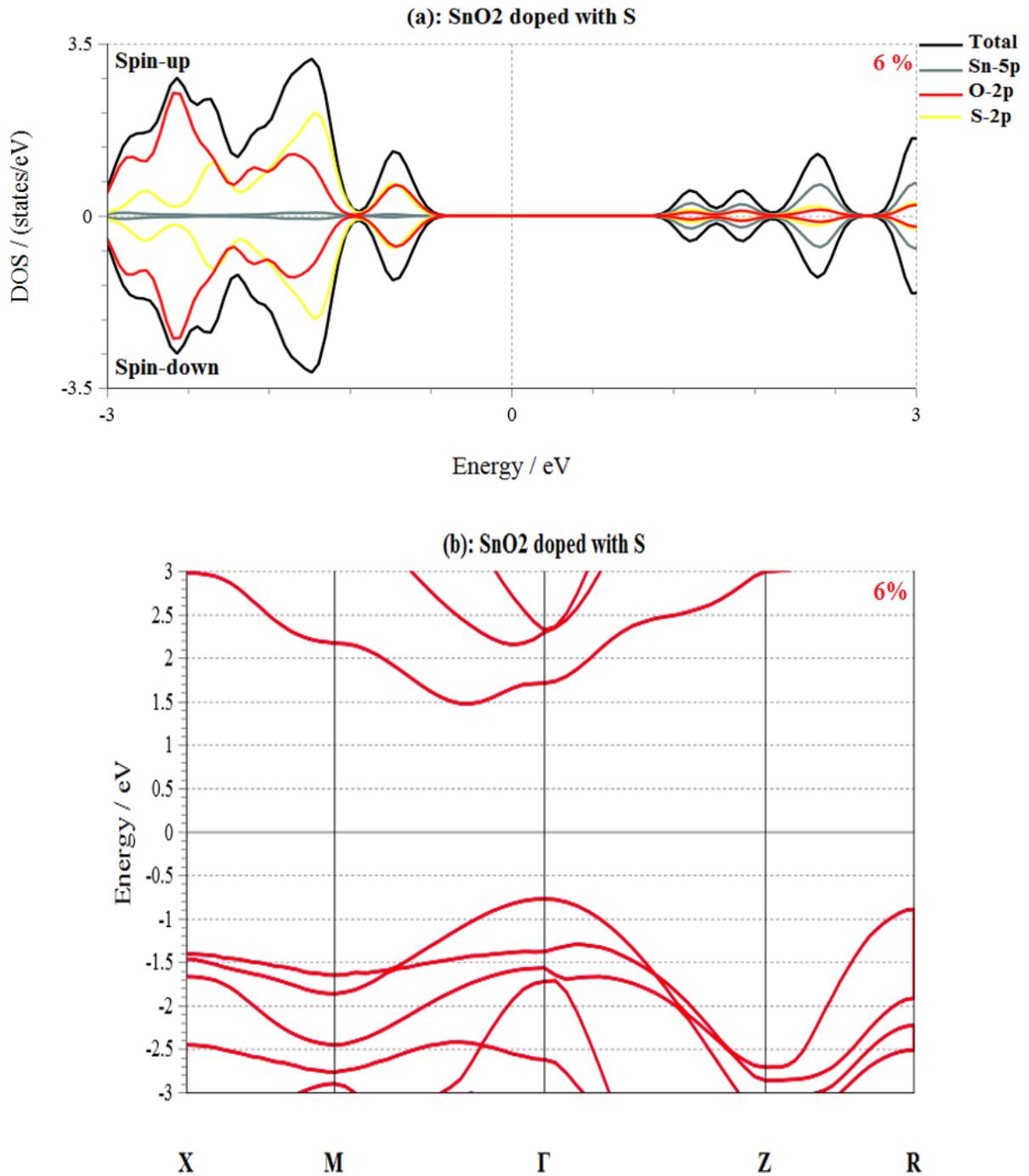

*Fig.4: Doped SnO2 material with 6% of Sulfur (S), Total and partial density of states (DOS) in (a). Band structure in (b).*

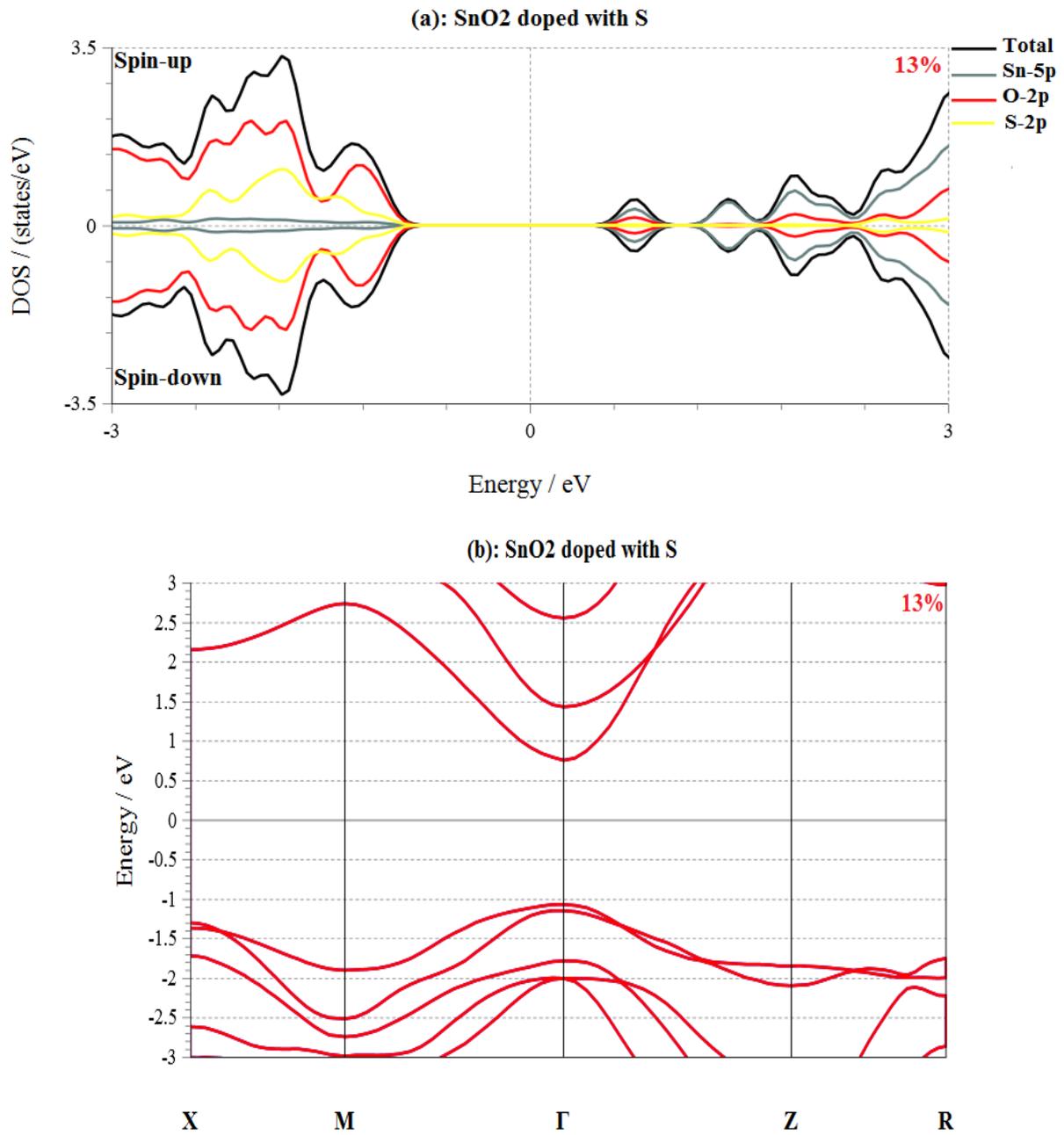

*Fig.5: Doped SnO2 material with 13 % of Sulfur (S), Total and partial density of states (DOS) in (a). Band structure in (b).*

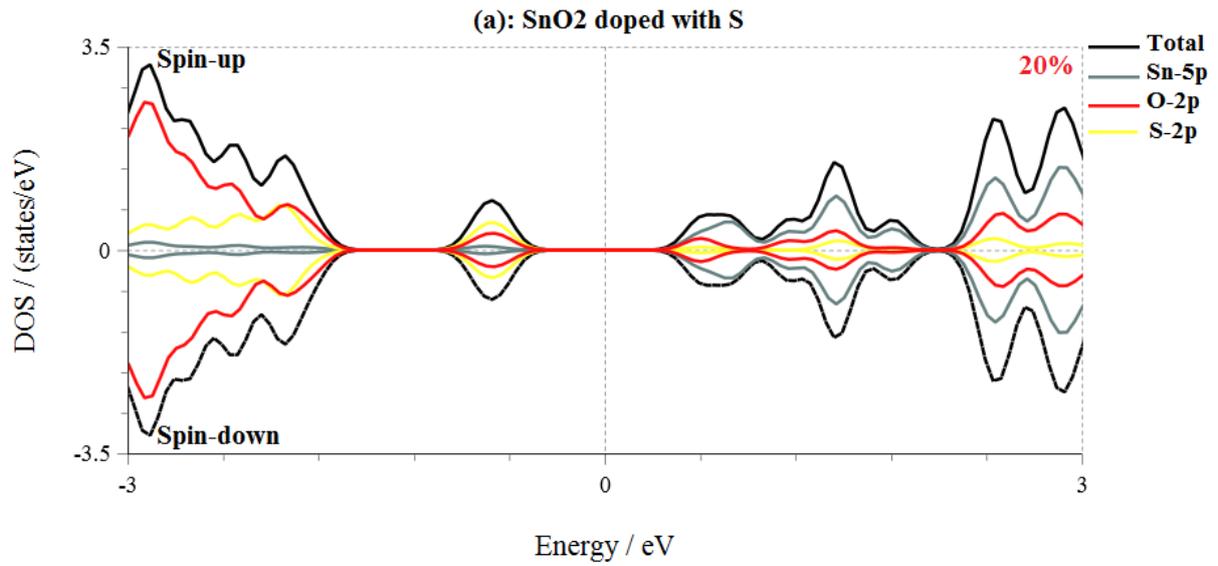

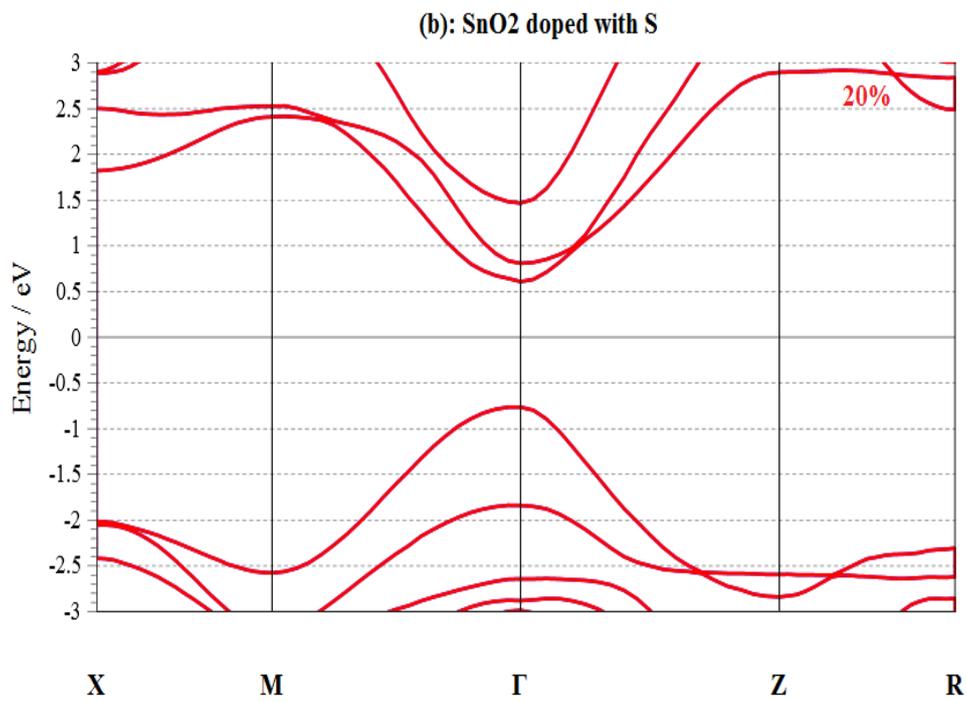

*Fig.6: Doped SnO2 material with 20% of Sulfur (S), Total and partial density of states (DOS) in (a). Band structure in (b).*

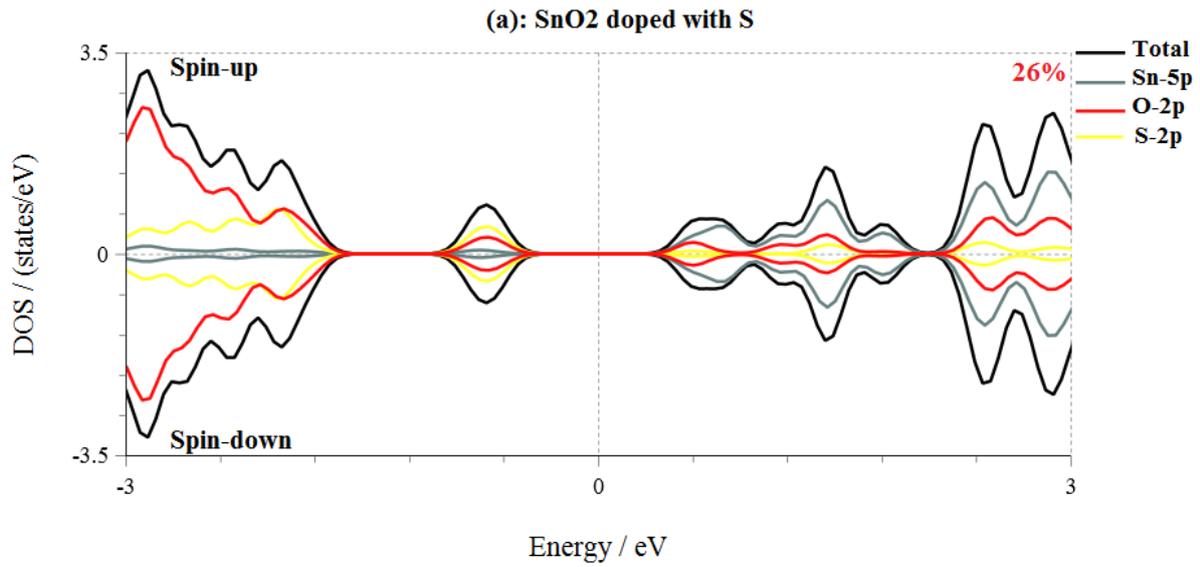

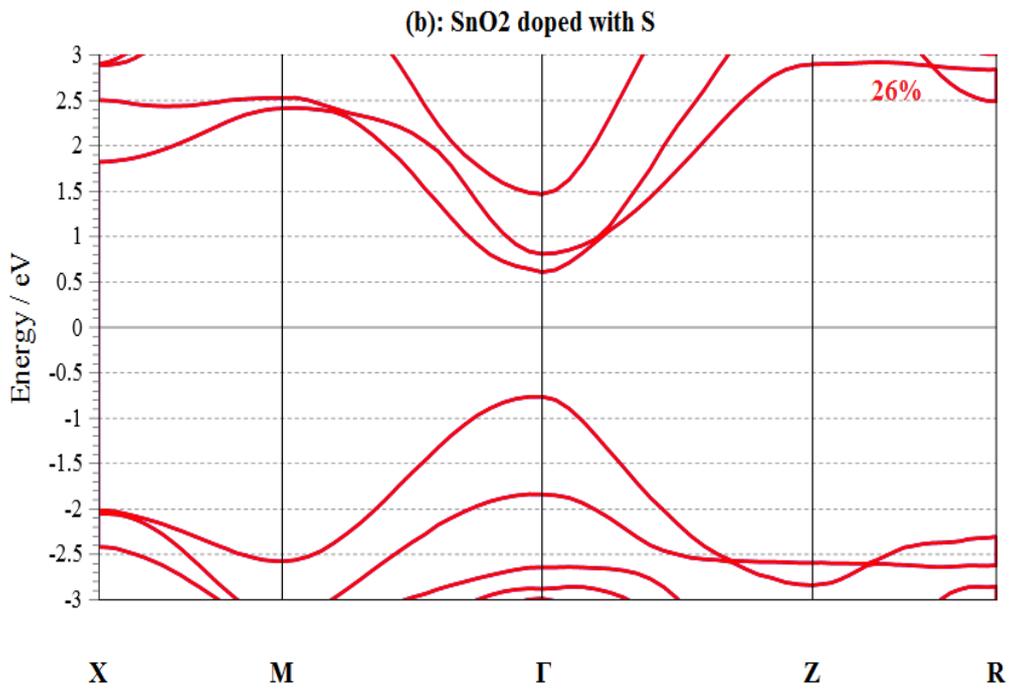

*Fig.7: Doped SnO2 material with 26% of Sulfur (S), Total and partial density of states (DOS) in (a). Band structure in (b).*

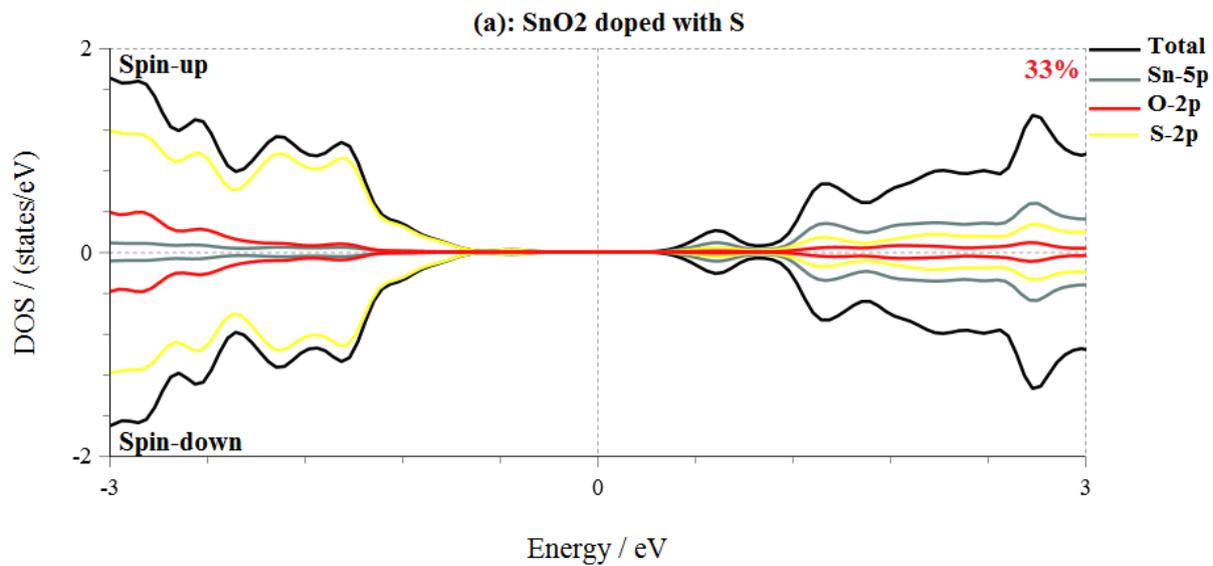

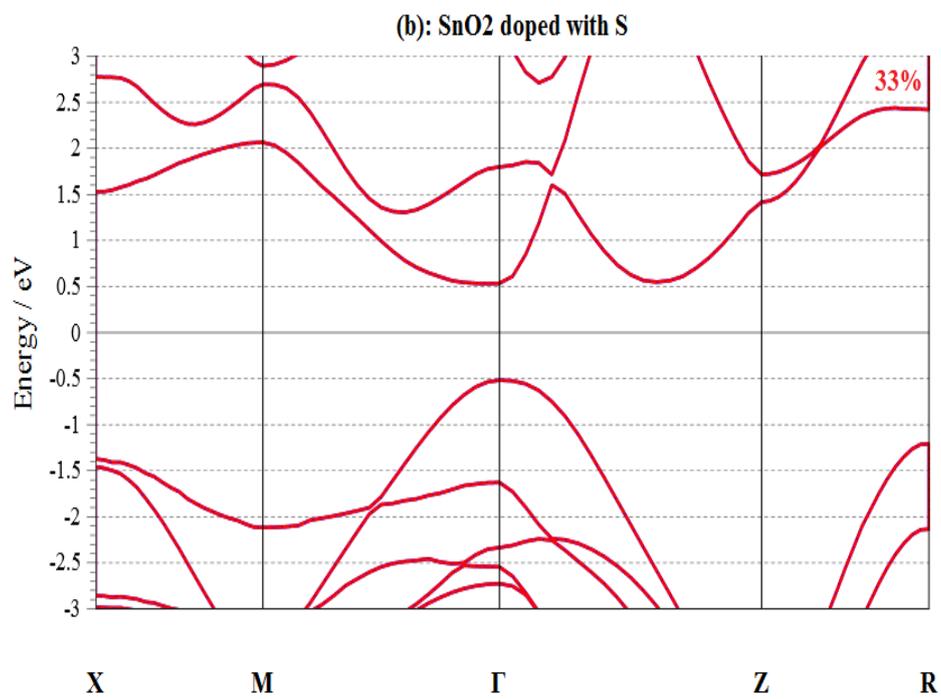

*Fig.8: Doped SnO2 material with 33% of Sulfur (S), Total and partial density of states (DOS) in (a). Band structure in (b).*

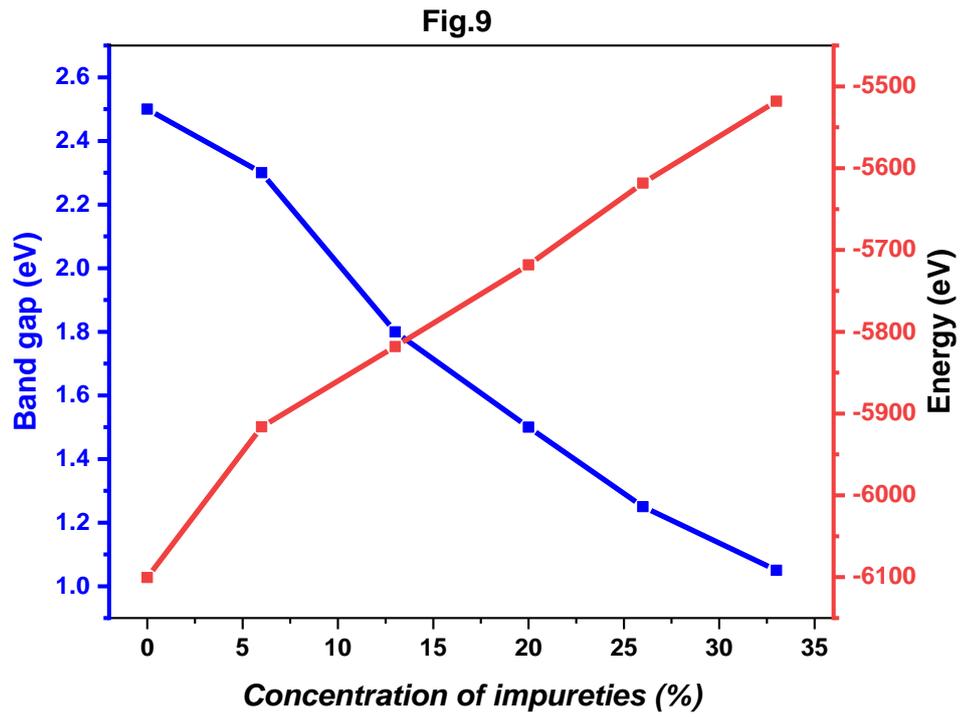

*Fig.9: Energy and band gap of the doped SnO2 material as a function of concentration (%) of impurities of sulfur.*

| Sulfur doped ZrO2 | Eg (eV) |
|---|---|
| Pure case | 2.5 present work<br>1.8 [60]<br>2.2 [61]<br>2.760 [62]<br>3.7 [63]<br>2.9 [64] |
| 6 % | 2.3 |
| 13 % | 1.8 |
| 20 % | 1.5 |
| 26 % | 1.25 |
| 33 % | 1.05 |

*Table 2: Band gap values of the pure and different concentration of the SnO2 doped material by sulfur.*

## IV.   Conclusion.

In summary, we have used the first-principle calculations which lead to results reported in the study of electronic properties of the SnO2 material doped by the sulfur atoms. In fact, we have used the ab-initio method applied on the basis of the Density Functional Theory (DFT) using the Quantum Espresso code. Indeed, the density of states and the band structure calculations for different concentrations have been illustrated.

It is found that when doping the SnO2 material with 6% of Sulfur (S), a perfect symmetry between up and down spin states in the total DOS is observed confirming the non-magnetic behavior of this material. It is also worth to note that the SnO2 material doped with 6% of Sulfur, exhibits a semiconductor of the P-type. On the other hand, the band gap decreases when increasing the concentration of doping the SnO2 by Sulfur. When comparing with the existing, our results are in good agreement with both the experimental and theoretical studies.

**References:**

[1] P. Pascariu Dorneanu, A. Airinei, M. Grigoras, N. Fifere, L. Sacarescu, N. Lupu, L. Stoleriu, Structural, optical and magnetic properties of Ni doped SnO2 nanoparticles, J. alloys and Compound, Vol. 668, (2016), 65-72 https://doi.org/10.1016/j. jallcom.2016.01.183.

[2] T. Akiyama, O. Wada, H. Kuwatsuka, T. Simoyama, Y. Nakata, K. Mukai, M. Sugawara, H. Ishikawa, Appl. Phys. Lett. 77 (2000) 1753.

[3] X. Duan, Y. Huang, R. Agarval, C.M. Lieber, Nature 421 (2003) 241.

[4] W. U. Wang, C. Chen, K. H. Lin, Y. Fang, C. M. Lieber, Proc. Natl. Acad. Sci. USA 102 (2005) 3208.

[5] M. H. Huang, S. Mao, H. Feick, H.Q. Yan, Y.Y. Wu, H. Kind, E. Weber, R. Russo, P.D. Yang, Science 292 (2001) 1897.

[6] D. J. Milliron, S.M. Hughes, Y. Cui, L. Manna, J. Li, L.W. Wang, A.P. Alivisatos, Nature 430 (2004) 190.

[7] H. Yu, J. Li, R.A. Loomis, L.W. Wang, W.E. Buhro, Nat. Mater. 2 (2003) 517.

[8] S. V. Kilina, C. F. Craig, D. S. Kilin, O. V. Prezhdo, J. Phys. Chem. C 111 (2007) 487.

[9] P. Pascariu (Dorneanu, A. Airinei, M. Grigoras, N. Fifere, L. Sacarescu, N. Lupu, L. Stoleriu, Structural, optical and magnetic properties of Ni doped SnO2 nanoparticles, Jour. alloys and Compou (2016), https://doi.org/10.1016/j. jallcom.2016.01.183.

[10] K. Sakthiraj, B. Karthikeyan, K. Balachandrakumar, Structural, optical and magnetic properties of copper (Cu) doped tin oxide (SnO2) nano crystal, Int.J. Chem. Tech Res. 7 (3) (2014-2015) 1481–1487.

[11] Z. Zhu, J. Zhou, H. Liu, Z. He, X. Wang, Graphene oxides as substrate for enhanced mammalian cell growth, J. Nanomater. Mol. Nanotechnol. 1 (2012) 2.

[12] S. Nilavazhagan, S. Muthukumaran, Investigation of optical and structural properties of Fe, Cu co-doped SnO2 nanoparticles, Superlattice. Microst. 83 (2015) 507–520.

[13] G. E. Patil, D. D. Kajale, S. D. Shinde, V. G. Wagh, V. B. Gaikwad, G. H. Jain, Synthesis of Cu-doped SnO2 thin films by spray pyrolysis for gas sensor application, Advancement in Sensing Technology (2013) 299–311.

[14] D. Xue, Z. Zhang, Y. Wang, Enhanced methane sensing performance of SnO2na-noflowers based sensors decorated with Au nanoparticles, Mater. Chem. Phys. 237(2019).

[15] N. Li, Y. Fan, Y. Shi, Q. Xiang, X. Wang, J. Xu, A low temperature formaldehyde gassensor based on hierarchical SnO/SnO2nano-flowers assembled from ultra thin nanosheets: synthesis, sensing performance and mechanism, Sens. Actuators B Chem. 294 (2019).

[16] J. Wang, W. Zeng, Z. Wang, Assembly of 2D nanosheets into 3Dflower-like NiO: synthesis and the influence of petal thickness on gas-sensing properties, Ceram. Int.42 (2016) 4567–4573.

[17] V. Inderan, M.M. Arafat, A.S.M.A. Haseeb, K. Sudesh, H.L. Lee, A comparative study of structural and ethanol gas sensing properties of pure, nickel and palladium dopedSnO2nanorods synthesised by the hydrothermal method, J. Phys. Sci. 30 (2019)127–143.

[18] F. Zhou, Z. Wang, B. Xu, L. Xia, X. Xiong, X. Sun, SnO2nanorod: an efficient non-noble-metal electrocatalyst for non-enzymatic H2O2sensing, Mater. Res. Express 6(2019).

[19] B. Jang, M.H. Kim, J. Baek, W. Kim, W. Lee, Highly sensitive hydrogen sensors: Pd-coated Si nanowire arrays for detection of dissolved hydrogen in oil, Sens. Actuators B Chem. 273 (2018) 809–814.

[20] L. Pan, Y. Zhang, F. Lu, Y. Du, Z. Lu, Y. Yang, T. Ye, Q. Liang, Y. Bando, X. Wang, Exposed facet engineering design of graphene-SnO2 nanorods for ultrastable Li-ion batteries, Energy Storage Mater. 19 (2019) 39–47.

[21] Y. Chen, T. Liu, C. Chen, W. Guo, R. Sun, S. Lv, M. Saito, S. Tsukimoto, Z. Wang, Synthesis and characterization of CeO2nano-rods, Ceram. Int. 39 (2013)6607–6610.

[22] M. J. Priya, P. M. Aswathy, M. K. Kavitha, M. K. Jayaraj, K. R. Kumar, Improved Acetone Sensing Properties of Electrospun Au-Doped SnO2Nanofibers, AIP Conference Proceedings, 2019.

[23] Z. Li, Q. Yang, Y. Wu, Y. He, J. Chen, J. Wang, La3+doped SnO2nanofibers forrapid and selective H2 sensor with long range linearity, Int. J. Hydrogen Energy 44 (2019) 8659–8668.

[24] W. Guo, M. Fu, C. Zhai, Z. Wang, Hydrothermal synthesis and gas-sensing proper-ties of ultrathin hexagonal ZnO nanosheets, Ceram. Int. 40 (2014) 2295–2298.

[25] D. Liu, J. Pan, J. Tang, W. Liu, S. Bai, R. Luo, Ag decorated SnO2 nanoparticles to enhance formaldehyde sensing properties, J. Phys. Chem. Solids 124 (2019) 36–43.

[26] G. Tofighi, D. Degler, B. Junker, S. Müller, H. Lichtenberg, W. Wang, U. Weimar, N. Barsan, J.-D. Grunwaldt, Microfluidically synthesized Au, Pd and Au Pd nano-particles supported on SnO2 for gas sensing applications, Sens. Actuators B Chem. 292 (2019) 48–56.

[27] D. Toloman, O. Pana, M. Stefan, A. Popa, C. Leostean, S. Macavei, D. Silipas, I. Perhaita, M.D. Lazar, L. Barbu-Tudoran, Photocatalytic activity of SnO2-TiO2 composite nanoparticles modified with PVP, J. Colloid Interface Sci. 542 (2019) 296–307.

[28] S. Zhang, C. Yin, L. Yang, Z. Zhang, Z. J. S. Han, A. B. Chemical, Investigation of the H2 sensing properties of multilayer mesoporous pure and Pd-doped SnO2 thin film, Sens. Actuators B Chem. 283 (2019) 399–406.

[29] T. Goto, T. Itoh, T. Akamatsu, N. Izu, W. J. S. Shin, A.B. Chemical, CO sensing properties of Au/SnO2–Co3O4 catalysts on a micro thermoelectric gas sensor, Sens. Actuators B Chem. 223 (2016) 774–783.

[30] J. Guo, J. Zhang, H. Gong, D. Ju, B. J. S. Cao, A. B. Chemical, Au nanoparticle-functionalized 3D SnO2 microstructures for high performance gas sensor, Sens. Actuators B Chem. 226 (2016) 266–272.

[31] Y. Wang, Z. Zhao, Y. Sun, P. Li, J. Ji, Y. Chen, W. Zhang, J. J. S. Hu, A. B. Chemical, Fabrication and gas sensing properties of Au-loaded SnO2 composite nano particles for highly sensitive hydrogen detection, Sens. Actuators B Chem. 240 (2017) 664–673.

[32] S. B. Kondawar, A. M. More, H. J. Sharma, S. P. Dongre, Ag-SnO2/Polyaniline composite nanofibers for low operating temperature hydrogen gas sensor, J. Mater. Nanosci. 4 (2017) 13–18 http://pubs.iscience.in/journal/index.php/jmns/article/view/632.

[33] Z. Lu, Q. Zhou, L. Xu, Y. Gui, Z. Zhao, C. Tang, W. J. M. Chen, Synthesis and characterization of highly sensitive hydrogen (H2) sensing device based on Ag doped SnO2 nanospheres, Materials 11 (2018) 492.

[34] A. M. Salih, Q. A. Drmosh, Z. H. Yamani, Silver nanoparticles decorated tin oxide thinfilms: synthesis, characterization, and hydrogen gas sensing, Front. Mater. 6 (2019) 188 https://www.frontiersin.org/articles/10.3389/fmats.2019.00188/full.

[35] S. Dhall, M. Kumar, M. Bhatnagar, B.R. Mehta, Dual gas sensing properties of graphene-Pd/SnO2 composites for H2and ethanol: role of nanoparticles-graphene interface, Int. J. Hydrogen Energy 43 (2018) 17921–17927.

[36] K. Suematsu, Y. Shin, N. Ma, T. Oyama, M. Sasaki, M. Yuasa, T. Kida, K. Shimanoe,Pulse-driven micro gas sensorfitted with clustered Pd/SnO2nanoparticles, Anal.Chem. 87 (2015) 8407–8415.

[37] Coey J M D, Douvails A P, Fitzgerald C B andVenkatesan M 2004Appl. Phys. Lett.841332

[38] P. Pascariu (Dorneanu, A. Airinei, M. Grigoras, N. Fifere, L. Sacarescu, N. Lupu, L. Stoleriu Structural, optical and magnetic properties of Ni doped $SnO_2$ nanoparticles, J. alloys and Compound, Vol. 668, (2016) Pages 65-72

[39] J. Divya, A. Pramothkumar, S. Joshua Gnanamuthu, D.C. Bernice Victoria, P.C. Jobe prabakar, Structural, optical, electrical and magnetic properties of Cu and Ni doped SnO2 nanoparticles prepared via Co-precipitation approach, Physica B: Condensed Matter, Volume 588, (2020), 412169, https://doi.org/10.1016/j.physb.2020.412169.

[40] S. IDRISSI, H. LABRIM, S. ZITI and L. BAHMAD , Investigation of the physical properties of the equiatomic quaternary Heusler alloys CoYCrZ (Z= Si and Ge) : A DFT study, journal of applied physics A 126(3), (2020) 190.

[41] S. Idrissi, H. Labrim, S. Ziti, L. Bahmad, Structural, electronic, magnetic properties and critical behavior of the equiatomic quaternary Heusler alloy CoFeTiSn, Physics Letters A, 2020, 126453, https://doi.org/10.1016/j.physleta.2020.126453.

[42] Idrissi, S., Labrim, H., Ziti, S. et al. Characterization of the Equiatomic Quaternary Heusler Alloy ZnCdRhMn: Structural, Electronic, and Magnetic Properties. J Supercond Nov Magn (2020). https://doi.org/10.1007/s10948-020-05561-8.

[43] S. IDRISSI, S. ZITI, H. LABRIM and L. BAHMAD, Critical magnetic behavior of the Rare Earth Based Alloy GdN: Monte Carlo simulations and DFT method, Accepted in the Journal of Materials Engineering and Performance 2020, (DOI: 10.1007/s11665-020-05214-w).

[44] S. Idrissi, S. Ziti, H. Labrim, L. Bahmad, Band gaps of the solar perovskites photovoltaic CsXCl3 (X=Sn, Pb or Ge), Materials Science in Semiconductor Processing, Volume 122, (2021) 105484, https://doi.org/10.1016/j.mssp.2020.105484.

[45] S. Idrissi, H. Labrim, L. Bahmad, A. Benyoussef, Study of the solar perovskite CsMBr3 (M=Pb or Ge) photovoltaic materials: Band-gap engineering, Solid State Sciences, Volume 118, 2021, 106679, https://doi.org/10.1016/j.solidstatesciences.2021.106679.

[46] S. Idrissi, S. Ziti, H. Labrim, L. Bahmad, Sulfur doping effect on the electronic properties of zirconium dioxide ZrO2, Materials Science and Engineering: B, Volume 270, 2021, 115200, https://doi.org/10.1016/j.mseb.2021.115200.

[47] S. Idrissi, H. Labrim, S. Ziti, L. Bahmad, A DFT study of the equiatomic quaternary Heusler alloys ZnCdXMn (X=Pd, Ni or Pt), Solid State Communications, Volume 331, 2021, 114292,
https://doi.org/10.1016/j.ssc.2021.114292.

[48] S. Idrissi, H. Labrim, L. Bahmad, A. Benyoussef, DFT and TDDFT studies of the new inorganic perovskite CsPbI3 for solar cell applications, Chemical Physics Letters, Volume 766, 2021, 138347, https://doi.org/10.1016/j.cplett.2021.138347. Idrissi, S., Labrim, H., Bahmad, L. et al. Structural, Electronic, and Magnetic Properties of the Rare Earth-Based Solar Perovskites: GdAlO3, DyAlO3, and HoAlO3. J Supercond Nov Magn (2021). https://doi.org/10.1007/s10948-021-05900-

[49] S. Idrissi, S. Ziti, H. Labrim, L. Bahmad, The critical magnetic behavior of the new Heusler CoXO2 alloys (X=Cu or Mn): Monte Carlo Study, Chinese Journal of Physics,
Volume 70, 2021, Pages 312-323, https://doi.org/10.1016/j.cjph.2021.01.008.

[50] P. Giannouzzi et al., J. Phys.: Condens. Matter 21, (2009), 395502; URL, ''http: //www.Quantum-espresso.org''.

[51] D. R. Hamann, M. Schlüter, C. Chiang Norm-conserving pseudopotentials Phys. Rev. Lett., 43 (1979), p. 1494.

[52] J. P. Perdew, J. A. Chevary, S. H. Vosko, K. A. Jackson, M. R. Pederson, D. J. Singh, C. Fiolhais, Atoms, molecules, solids, and surfaces: applications of the generalized gradient approximation for exchange and correlation, Phys. Rev. B, 46, (1992), 6671-6687.

[53] H. J. Monkhorst, J.D. Pack Phys. Rev. B, 13 (1976), p. 5188

[54] I. Phys. Chem. Solids Vol. 48, No. 2, pp. 171-180, 1987

[55] K. Momma and F. Izumi, "VESTA 3 for three-dimensional visualization of crystal, volumetric and morphology data," J. Appl. Crystallogr., 44, 1272-1276 (2011).

[56] Michel Levy, Thierry Pagnier, Ab initio DFT computation of SnO2 and WO3 slabs and gas–surface interactions, Sensors and Actuators B: Chemical, Volume 126, Issue 1, 2007, Pages 204-208, https://doi.org/10.1016/j.snb.2006.11.047.

[57] F. E. H. Hassan, A. Alaeddine, M. Zoaeter, I. Rachidi, Int. J. Mod. Phys. B, 19 (2005), p. 4081

[58] L. Gracia, A. Beltrán, J. Andrés, J. Phys. Chem. B 111 (2007) 6479.

[59] A. M. Mazzone, Phys. Rev. B 68 (2003) 045412.

[60] V. Stambouli, A. Zebda, E. Appert, C. Guiducci, M. Labeau, J.P. Diard, B.L. Gorrec, N. Brack, P. J. Pigram Electrochim. Acta, 51 (2006), p. 5206

[61] Jian Xu, Shuiping Huang, Zhanshan Wang, First principle study on the electronic structure of fluorine-doped SnO2, Solid State Communications, Volume 149, Issues 13–14, 2009, Pages 527-531, https://doi.org/10.1016/j.ssc.2009.01.010.

[62] F. El Haj Hassan, S. Moussawi, W. Noun, C. Salameh, A.V. Postnikov, Theoretical calculations of the high-pressure phases of SnO2, Computational Materials Science, Volume 72, 2013, Pages 86-92, https://doi.org/10.1016/j.commatsci.2013.02.011.

[63] D. Sanchez-Porta, E. Artacho, J.M. SolerSolid State Commun., 95 (1995), p. 685

[64] A. Ferreira da Silva, I. Pepe, C. Person, J. Souza de Almeida, C. Moysés Araújo, B. Johansson, C. Y. An, J. H. Guo, Phys. Scr. T109 (2004) 180.

[65] L. Soussi, T. Garmim, O. Karzazi, A. Rmili, A. El Bachiri, A. Louardi, H. Erguig, Effect of (Co, Fe, Ni) doping on structural, optical and electrical properties of sprayed SnO2 thin film, Surfaces and Interfaces, Volume 19, (2020), 100467, https://doi.org/10.1016/j.surfin.2020.100467.